\documentclass[fleqn,usenatbib]{mnras}

\usepackage{newtxtext,newtxmath}

\usepackage[T1]{fontenc}
\usepackage{float}

\DeclareRobustCommand{\VAN}[3]{#2}
\let\VANthebibliography\thebibliography
\def\thebibliography{\DeclareRobustCommand{\VAN}[3]{##3}\VANthebibliography}

\usepackage{graphicx}	
\usepackage{amsmath}
\usepackage{orcidlink}
\usepackage{subcaption}
\usepackage{booktabs}
\usepackage{ragged2e}

\title[Multiwavelength follow-up of EP241107a]{ Characterising EP241107a: Multiwavelength Observations of an\textit{ Einstein Probe}-detected Fast X-ray Transient}

\def\EPb{EP241107a}
\def\flux{erg~cm$^{2}$ s$^{-1}$}
\def\lum{erg~s$^{-1}$}
\def\msun{M_\odot}

\def\arcsec{\hbox{$^{\prime\prime}$}}
\def\approxlt{\ifmmode \rlap{$<$}{}_{{}_{{}_{\textstyle\sim}}} \else%
$\rlap{$<$}{}_{{}_{{}_{\textstyle\sim}}}$\fi}

\author[Eappachen et al.]{
D. Eappachen\orcidlink{0000-0001-7841-0294}$^{1}$\thanks{E-mail: deepak.eappachen@iiap.res.in},
A. Balasubramanian\orcidlink{0000-0003-0477-7645} $^{1}$,
Vishwajeet Swain\orcidlink{0000-0002-7942-8477} $^{2}$,
G. C. Anupama \orcidlink{0000-0003-3533-7183}$^{1}$,
D. K. Sahu \orcidlink{0000-0002-6688-0800}$^{1}$,
\newauthor
V. Bhalerao \orcidlink{0000-0002-6112-7609}$^{2}$,
T. Ahumada \orcidlink{0000-0002-2184-6430} $^3$,
I. Andreoni \orcidlink{0000-0002-8977-1498} $^{4}$,
Sudhanshu Barway \orcidlink{0000-0002-3927-5402}$^{1}$,
J. Carney \orcidlink{0000-0001-8544-584X
}$^{4}$, 
J. Freeburn \orcidlink{0009-0006-7990-0547}  $^{4}$,
\newauthor
M. M. Kasliwal \orcidlink{0000-0002-5619-4938} $^{3}$,
Tanishk Mohan \orcidlink{0009-0001-4683-388X}$^{2}$,
A. C. Rodriguez \orcidlink{0000-0003-4189-9668} $^{5}$,
G. Waratkar \orcidlink{0000-0003-3630-9440} $^{3,2}$
\\
\\
$^{1}$Indian Institute of Astrophysics, II Block, Koramangala, Bengaluru, Karnataka, India, 560034\\
$^{2}$ Department of Physics, Indian Institute of Technology Bombay, Powai, 400 076, India\\
$^{3}$Cahill Center for Astrophysics, California Institute of Technology, MC 249-17, 1216 E California Boulevard, Pasadena, CA, 91125, USA \\
$^{4}$ Department of Physics and Astronomy, University of North Carolina at Chapel Hill, Chapel Hill, NC 27599-3255, USA \\
$^{5}$Center for Astrophysics | Harvard \& Smithsonian, 60 Garden Street, Cambridge, MA 02138, USA\\
}

\date{Accepted XXX. Received YYY; in original form ZZZ}

\pubyear{\the\year{}}

\begin{document}
\label{firstpage}
\pagerange{\pageref{firstpage}--\pageref{lastpage}}
\maketitle

 \begin{abstract}
Fast X-ray Transients (FXTs) represent a new class of highly luminous transients in soft X-rays ($\sim$0.3--10 keV) associated with violent astrophysical processes. They manifest as short, singular flashes of X-ray photons with durations lasting from minutes to hours. Their origin remains unclear, and they have been associated with various progenitor mechanisms. 
The newly launched X-ray survey, {\it Einstein-Probe} (EP), is revolutionising this field by enabling the discovery and immediate follow-up of FXTs.
Here we present the multiwavelength observations of EP--discovered FXT  EP241107a and the discovery of its radio counterpart. Comparison of the optical and radio observations of EP241107a and its host properties with other extragalactic transients suggests a gamma-ray burst (GRB) origin. 
Through our afterglow modelling, we infer the GRB jet properties for EP241107a, yielding a jet of the isotropic-equivalent kinetic energy $E_{\mathrm{K,iso}} \sim10^{51}$ erg, with a half opening
angle $\theta_{c}$ $\approx$15$^{\circ}$, viewed at an angle of $\theta_{\rm obs}$~$\approx$9$^{\circ}$.
We also evaluate EP241107a in the landscape of both EP-discovered FXTs as well as the FXTs discovered from \textit{Chandra}, \textit{XMM}-Newton and \textit{Swift}-XRT.
\end{abstract}




\section{Introduction}
Fast X-ray transients (FXTs) are bursts in soft X-rays ($\sim$0.3--10~keV) with durations ranging from a few hundred to 10$^4$ s. Before the launch of {\it Einstein-Probe} (EP; \citealt{2022hxga.book...86Y}), roughly 95 per cent of the FXTs were discovered through searches in X-ray archival data, especially from {\it Chandra} (\citealt{2013jonker}; \citealt{glennie}; \citealt{}\citealt{Bauer}; \citealt{Xue2019}; \citealt{Quirola2022}; \citealt{2022ApJ...927..211L}; \citealt{Quirola2023}; \citealt{2023ApJ...948...91E}) and  XMM-{\it Newton} archives (\citealt{Alp2020}; \citealt{Novara2020}; \citealt{2020A&A...640A.124P}). Most FXTs were identified and reported months or years after the bursts, thereby lacking prompt follow-ups and, therefore, no multiwavelength counterparts were detected except in one case. It was in the case of SN2008D, where a serendipitous detection of the FXT associated with a supernova shock breakout (SN SBO) was reported by \citet{2008Natur.453..469S} using {\it Swift}-XRT. In the absence of multiwavelength counterparts, significant efforts have been made to study the associated host galaxies of the FXTs, thereby constraining their origin (e.g., \citealt{Xue2019}; \citealt{2022ApJ...927..211L}; \citealt{2022MNRAS.514..302E}; \citealt{2024MNRAS.52711823E}; \citealt{2024arXiv240707257I}; \citealt{anneetal}; \citealt{2024arXiv241010015Q}).


FXTs represent a heterogeneous and still poorly understood class of X-ray transients whose diversity implies that no single progenitor model can adequately account for their origin.
Major progenitor mechanisms include i) an SN SBO where an X-ray flash is produced as the shock wave from a supernova crosses the surface of the star (\citealt{2008Natur.453..469S}; \citealt{2017hsn..book..967W}; \citealt{2022ApJ...933..164G}) ii) a white dwarf (WD) --intermediate-mass black hole (IMBH) tidal disruption event (TDE) which could produce bursts in X-rays that last for shorter timescales compared to supermassive black hole TDE (\citealt{2016ApJ...819....3M}; \citealt{2020SSRv..216...39M}) iii) a long gamma-ray burst (lGRB) association where X-ray emission is produced when a mildly relativistic cocoon jet breaks the surface of a massive progenitor star (\citealt{2002MNRAS.337.1349R}; \citealt{2017ApJ...834...28N}) or iv) an X-ray emission from a magnetar formed as a result of a binary neutron star merger (\citealt{2014MNRAS.439.3916M}; \citealt{2017ApJ...835....7S}). \citet{2024A&A...690A.101W} { suggest a region of parameter space where the observational properties of off-axis GRB afterglows align with those of FXTs}.

EP is revolutionising the field of FXTs by discovering and reporting X-ray transients in the X-ray band of 0.5--4 keV within a few minutes/hours of the FXTs, enabling prompt follow-up. {Follow-up of some of these FXTs reveals an extragalactic nature:} for instance, EP240315a (\citealt{2024ApJ...969L..14G}; \citealt{2024arXiv240416350L}; \citealt{2024arXiv240416425L}) at a redshift of $z$= 4.859 is consistent with a lGRB (\citealt{2024ApJ...969L..14G}; \citealt{2024arXiv240416350L}; \citealt{2024arXiv240718311R}). Recent multiwavelength observations of the EP-FXTs provide further evidence supporting multiple origins (\citealt{2024ApJ...969L..14G}; \citealt{2024arXiv240416350L}; \citealt{2024arXiv240919070S}; \citealt{2024arXiv240919056V}; \citealt{2025ApJ...988L..13R}; \citealt{2025arXiv250417516S}; \citealt{2025arXiv250408886E}; \citealt{2025arXiv250714286L}; \citealt{2025arXiv250813039J}).

 One of the well-studied EP-discovered FXTs is the recently reported EP240414a (\citealt{2024arXiv240919056V}; \citealt{2024arXiv240919070S}; \citealt{2024arXiv241002315S}; \citealt{2024arXiv240919055B}) which was found to be located in a massive galaxy at a redshift of $z$~=~0.401 at a projected distance of 25.7 kpc from the galactic centre. The optical light curve of the counterpart 
 shows three different components which are suggested to be due to the initial dominant contribution from a thermal cocoon emission, followed by the circumstellar medium-interaction and the third phase dominated by the SN radioactive decay (\citealt{2024arXiv240919056V}). The fast rise time in the second phase of the light curve of the optical counterpart of EP240414a along with the extremely blue optical spectrum is similar to properties seen in Luminous Fast Blue Optical Transients (LFBOTs; \citealt{2018ApJ...865L...3P}). At later time scales, the spectrum shows similar features to that of broad-lined Type Ic SNe, suggesting a collapsar origin, although no gamma-ray counterpart has been detected. Another EP-FXT EP241021a (\citealt{2025arXiv250314588B}; \citealt{2025arXiv250505444G}; \citealt{2025arXiv250507665X}; \citealt{2025arXiv250508781Y}; \citealt{2025arXiv250512491W}) also shows a very similar optical afterglow light curve to that of EP240414a. However, for EP241021a, no features of an associated  SN were found. The detection of the radio counterpart and the apparent absence of a gamma-ray counterpart indicate that a low luminosity GRB (LL-GRB) could have gone undetected or an off-axis or {choked} jet could be responsible for the radio emission. In addition to both the above-discussed EP-FXTs, an intensive follow-up campaign to search for the counterpart of EP240408a was reported (\citealt{2025ApJ...979L..30O}). However, deep optical and radio observations resulted in non-detections of the transient. \citet{2025ApJ...979L..30O} favour a peculiar GRB or jetted WD-IMBH TDE at high redshift ($z \gtrsim 1$), though neither perfectly explains the observations.
 This highlights the diverse range of transients that EP is unveiling, including FXTs with distinct origins. {The redshift distribution of these EP-discovered FXTs offers valuable insights into their possible progenitor mechanisms} (\citealt{2025arXiv250907141O}).

In this paper, we report our multiwavelength follow-up observations of the FXT \EPb{}. The  FXT \EPb{} was detected by the Wide-field X-ray Telescope (WXT) on board
the EP,  at 2024-11-07T14:10:23 UTC (\citealt{GCN38112}).  \EPb{} trigger flux is estimated to be around 10$^{-10}$ \flux in 0.5$-$4 keV band. The follow-up X-ray telescope (FoXT) \footnote{EP follow-up X-ray telescope is conventionally named EP-FXT. However,
to avoid confusion with fast X-ray transients (FXT), we are using the EP-FoXT
acronym for the EP follow-up X-ray telescope} on EP carried out follow-up observations about five minutes later, which detected an X-ray source at R.A. = $02^{\rm h}$$20^{\rm m}$$02.04^{\rm s}$, Dec = $+03^{\rm \circ}$ $19^{\rm \prime}$ $58.44^{\rm \prime \prime}$ with an uncertainty of about 10\arcsec. The follow-up observations from the Al-Khatim Observatory detected and reported the detection of an optical counterpart at  R.A. and Dec $02^{\rm h}$$20^{\rm m}$$02.45^{\rm s}$, $+03^{\rm \circ}$ $20^{\rm \prime}$ $02.2^{\rm \prime \prime}$  in the $I_c$-band at a magnitude of 17.85$\pm$0.18 (\citealt{GCN38115}). Though reported later, the earliest optical counterpart detection for \EPb{} was by \citet{} Chinese Ground Follow-up Telescope of \emph{SVOM} mission, which detected the counterpart a magnitude of 17.12 $\pm$ 0.10 in $i$-band, starting the observations just $\sim$ 5.5 minutes after the EP-WXT trigger (\citealt{GCN38116}).

 {We use our multiwavelength observations of the transient counterpart, along with publicly available data, combined with our afterglow and host modelling, to derive constraints on the nature of}  \EPb{}. Throughout the paper; we assume a flat $\Lambda$-CDM cosmology, with Hubble constant  H$_\mathrm{0}$= 67.8 $\pm$ 0.9 km s$^{-1}$Mpc$^{-1}$ and matter density parameter $\Omega_m$=0.308 $\pm$  0.012 (\citealt{2016A&A...594A..13P}). Magnitudes are quoted in the AB system.

\label{intro}

\begin{figure*}
\centering
\includegraphics[scale=1]{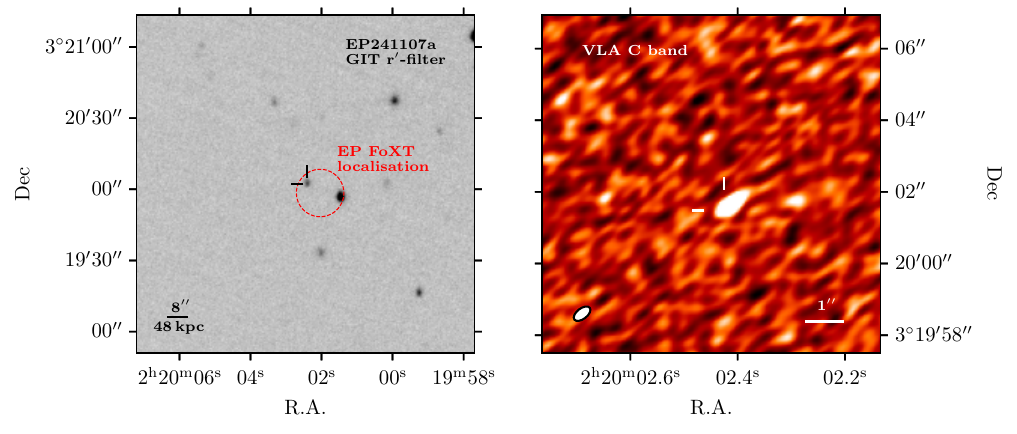}
\caption{The location of the counterparts to \EPb{} in optical and radio bands is shown. The {\it left panel} shows the GIT $r{^\prime}$-filter image of the field of \EPb{}, 0.165 days after the EP-trigger. The EP-FoXT uncertainty region of 10\arcsec{} is marked with a dashed red circle, while the optical counterpart of \EPb{} is marked with black lines. The VLA C-band image of the field of \EPb{} is shown in the {\it right panel}. The VLA beam size for this observation is given in the bottom left of the image. Note that the scales of the optical and radio images are different. } 
\label{detection_plot}
\end{figure*}

\section{Observations and Analysis }

\begin{table*}
\small
\begin{center}
\caption{A journal of optical and radio observations of \EPb{} used in this paper.}
\label{tab}

\subcaption{The uGMRT and VLA radio observations of \EPb{}}
\begin{tabular}{ccccc}
\hline
\multicolumn{1}{|p{2.5cm}|}{\centering Date and Time (UT)}
&\multicolumn{1}{|p{0.5cm}|}{\centering $\Delta$T}
&\multicolumn{1}{|p{1.5cm}|}{\centering Telescope}
&\multicolumn{1}{|p{1 cm}|}{\centering Frequency (MHz)}
&\multicolumn{1}{|p{1.5cm}|}{\centering Flux density ($\mu$Jy)}\\
\hline
2024-11-24T00:54:58.0 & 16.45 & VLA & 10000 & 232 $\pm$ 9\\
2024-12-01T00:38:08.0 & 23.44 & VLA & 6000 & 207 $\pm$ 8\\
2024-12-17T16:39:48.6 & 40.10 & uGMRT & 1265 & $<72$\\
2024-12-20T14:43:24.2 & 43.02 & uGMRT & 648 & $<255$\\

\hline
\end{tabular}

\caption{Optical photometric observations of \EPb{}}
\begin{tabular}{lcccccc}
\hline
UT (start) & $\Delta$T (days) & Telescope/Instrument & 
 Optical Band & Magnitude & Error & Reference \\
\hline
2024-11-07 00:05:45.600 & 0.004 & Chinese Ground Follow-up Telescope & i & 17.12 & 0.10 & \citet{GCN38116} \\
2024-11-07 01:43:40.800 & 0.072 & Al-Khatim Observatory & Ic & 17.85 & 0.18 & \citet{GCN38128} \\
2024-11-07 01:45:07.200 & 0.073 & Al-Khatim Observatory & Ic & 17.95 & 0.19 & \citet{GCN38128}  \\
2024-11-07 03:13:40.800 & 0.095 & Al-Khatim Observatory & Ic & 18.11 & 0.24 & \citet{GCN38128}  \\
2024-11-07 04:41:25.600 & 0.116 & Al-Khatim Observatory & Ic & 18.48 & 0.25 & \citet{GCN38128}  \\
2024-11-07 16:07:26.667 & 0.165 & GIT & r & 19.45 & 0.22 & This work \\
2024-11-07 18:23:24.333 & 0.176 & GIT & g & 19.99 & 0.05 & This work \\
2024-11-07 18:39:20.667 & 0.187 & GIT & i & 19.17 & 0.06 & This work \\
2024-11-07 18:49:55.200 & 0.210 & 1-m Las Cumbres Observatory & i & 19.40 & 0.10 &  \citet{2024GCN.38127....1L}\\
2024-11-07 22:02:52.000 & 0.263 & Fraunhofer Telescope Wendelstein & r & 19.65 & 0.01 &  \citet{GCN38120}\\
2024-11-07 22:02:52.000 & 0.263 & Fraunhofer Telescope Wendelstein & i & 19.40 & 0.01 &  \citet{GCN38120}\\
2024-11-07 22:02:52.000 & 0.263 & Fraunhofer Telescope Wendelstein & J & 18.71 & 0.01 &  \citet{GCN38120}\\
2024-11-07 22:41:42.667 & 0.295 & 1.93m OHP & r & 19.80 & 0.05 & \citet{GCN38122}  \\
2024-11-07 21:28:45.000 & 0.304 & GIT & r & 19.68 & 0.21 & This work \\
-- & 0.67 & KAIT & R & 20.50 & 0.1 &  \citet{2024GCN.38136....1Z}\\
2024-11-08 14:13:50.667 & 1.002 & GIT & i & >20.36 & -- & This work \\
2024-11-08 14:45:12.667 & 1.024 & GIT & g & >21.68 & -- & This work \\
2024-11-08 14:52:57.556 & 1.030 & GIT & r & 21.59 & 0.25 & This work \\
2024-11-08 17:53:52.000 & 1.024 & 1m LOT Lulin Observatory & r & 21.50 & 0.32 &  \citet{2024GCN.38131....1K} \\
2024-11-08 18:51:56.750 & 1.196 & GIT & r & 21.51 & 0.25 & This work \\
2024-11-08 20:47:08.583 & 1.276 & GIT & i & >21.16 & -- & This work \\
2024-11-09 20:54:24.031 & 2.281 & HCT/HFOSC & r & >22.5 & -- & This work \\
2024-11-10 15:30:05.667 & 3.055 & GIT & r & >21.43 & -- & This work \\
2024-11-10 16:07:51.000 & 3.082 & GIT & i & >20.92 & -- & This work \\
2024-11-11 03:05:18.000 & 3.480 & SOAR/Goodman & g & >22.40 & -- & This work \\
2024-11-11 03:24:00.000 & 3.490 & SOAR/Goodman & r & >22.90 & -- & This work \\
2024-11-11 03:38:26.000 & 3.510 & SOAR/Goodman & i & >22.70 & -- & This work \\
2024-11-12 16:02:47.600 & 5.078 & GIT & r & >20.86 & -- & This work \\
2024-11-12 16:29:24.200 & 5.097 & GIT & i & >20.56 & -- & This work \\
\hline
\hline
\end{tabular}
\end{center}
\end{table*}

\subsection{GROWTH-India Telescope}
The GROWTH-India Telescope (GIT; \citealt{2022AJ....164...90K}), located at the Indian Astronomical Observatory (IAO), Hanle, India, is a 0.7~m robotic telescope with a 0.7$^{\circ}$ field of view and is dedicated to observing the transient sky. We observed the field of \EPb{} in Sloan $g{^\prime}$, $r{^\prime}$ and $i{^\prime}$ filters within $\sim$3.55 hours after EP-WXT trigger. We utilised a \textsc{python}-based pipeline (\citealt{2022AJ....164...90K}) to obtain the photometry, making use of the Pan-STARRS DR2 (\citealt{2016arXiv161205560C})  catalogue for photometric and astrometric calibration.

An optical counterpart was detected at an $r{^\prime}$-filter magnitude of 19.42 $\pm$ 0.07 at the same R.A and Dec as reported by \citealt{GCN38128}. The GIT detection image of \EPb{} is shown in {\it left}-panel of Fig. \ref{detection_plot}.  We obtained the photometry of the counterpart to \EPb{} in two more epochs, and the magnitudes and upper limits are given in Table \ref{tab}.

\subsection{Himalayan-Chandra Telescope}
We observed the field of \EPb{} using the  Himalaya Faint Object Spectrograph Camera (HFOSC) mounted at the 2~m Himalayan Chandra Telescope (HCT) of the Indian Astronomical Observatory (IAO), Hanle, India. The observations started at 2024-11-09T20:54:24.03 UT in  $r{^\prime}$-filter.  We used \textsc{iraf} (\citealt{1986SPIE..627..733T}) for bias and flat field correction, whereas for cosmic removal, L.A. Cosmic (\citealt{2001PASP..113.1420V}) was employed. A total of 540-second exposure was obtained, and we used the \textsc{ iraf imcombine} task for stacking. We applied the astrometric correction to the stacked image using astrometry.net\footnote{https://nova.astrometry.net/upload} (\citealt{2010AJ....139.1782L})

At $\sim$2.28 days after the trigger, we did not find any associated counterpart for \EPb{} in the HCT $r{^\prime}$-image. The photometric calibration of the image was done using stars from the Pan-STARRS DR2
catalogue data (\citealt{2016arXiv161205560C}). We deduced a 5$\sigma$ optical upper limit of 22.5 in $r{^\prime}$-filter 2.28 days after the EP-trigger.

\subsection{The 4.1-meter Southern Astrophysical Research Telescope}
\begin{figure*}
\centering
\includegraphics[scale=0.65]{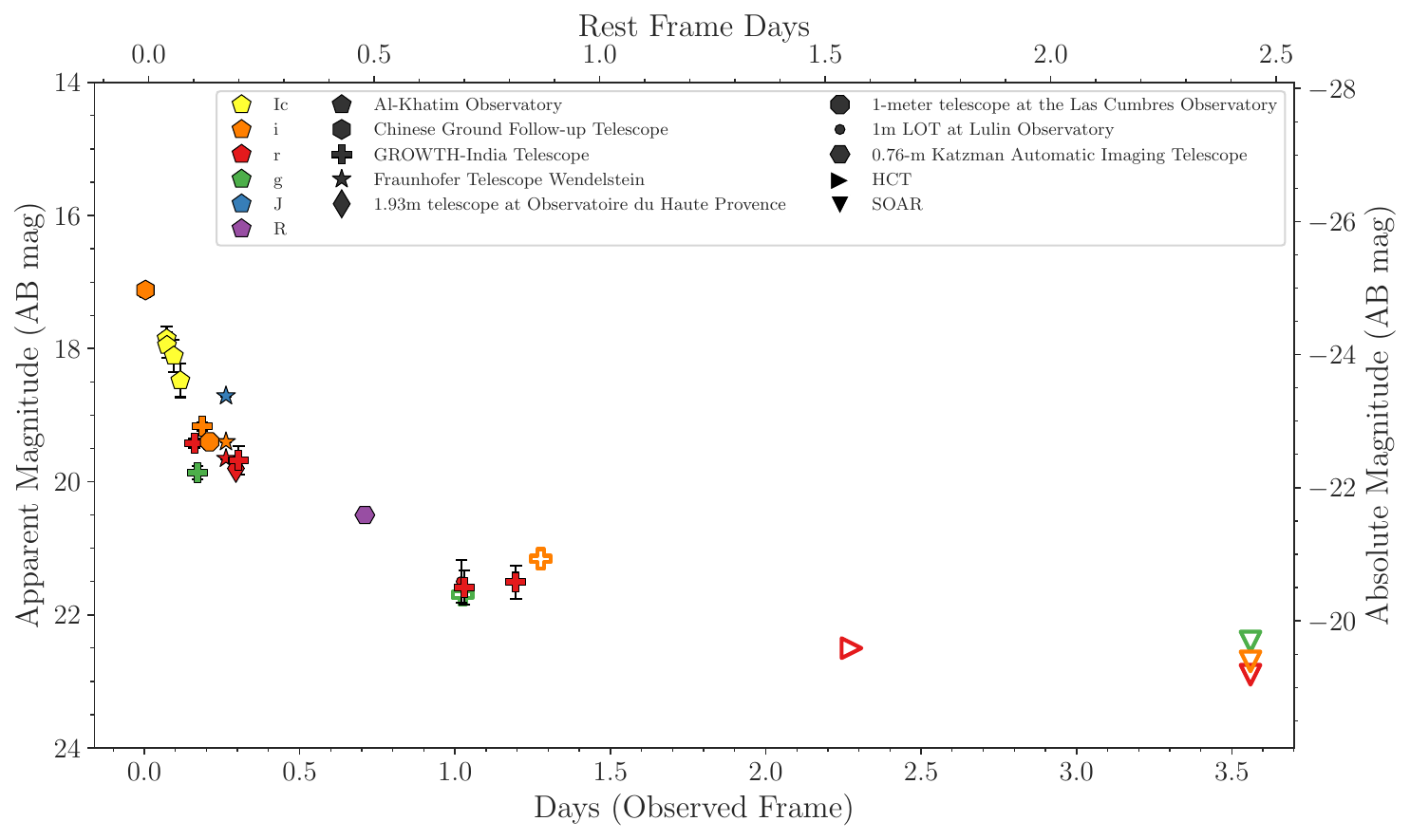}
\caption{Optical photometry of the counterpart to \EPb{}. We present the data obtained from our follow-up observations along with the data obtained from the GCNs. Different telescopes are denoted with different markers, while filters are indicated by various colours. The upper limits obtained from our observations are represented by open markers. We consider the T$_0$ to be the time of the EP-WXT trigger. Chinese Ground Follow-up Telescope at Changchun Observatory started the observation of the field of \EPb{} about six minutes after the trigger and detected the counterpart in $i$-filter at a magnitude of 17.12$\pm$0.10. {The secondary x and y axes assume a redshift of $z$ = 0.457$\pm$0.003.} } 
\label{fig:ep241107a_lc}
\end{figure*}

With the Goodman High Throughput Spectrograph's red camera, mounted on the Southern Astrophysical Research (SOAR; \citealt{2004SPIE.5492..331C}) telescope, the location of the optical counterpart to EP241107a was observed \citep[PI: Andreoni;][]{2024GCN.38208....1F} $\sim$3.5\,d after the X-ray transient.  Difference imaging was conducted with SFFT \citep{2022ApJ...936..157H}, using templates from the DECam Legacy Survey \citep{2019AJ....157..168D}.  The optical counterpart was not detected in these observations with 5$\sigma$ upper limits of $g=22.4$, $r=22.9$ and $i=22.7$. The photometric light curve of \EPb{} is shown in Fig.~\ref{fig:ep241107a_lc}.

\subsection{Keck LRIS}
Spectroscopic observation of \EPb{} was carried out on the Keck~I telescope, starting at 2024-11-08T08:34:17.9 UT using the Low-Resolution Imaging Spectrometer (LRIS; \citealt{1995PASP..107..375O}). Using the 600 $\ell$ 
mm$^{-1}$ blue grism ($\lambda_{\rm blaze} =$
4000 \AA), the 400 $\ell$ mm$^{-1}$ red grating ($\lambda_{\rm blaze} =$ 8500 \AA), the 5600 \AA\, dichroic, and the 1 arcsec slit we obtained a 900 second observation covering the full optical window at moderate resolving power, $R \equiv \lambda / \Delta \lambda \approx 1500$ for objects filling the slit.   We used BD+284211 observed with the same instrument configuration for flux calibration. We used the code Lpipe to reduce and flux calibrate the spectrum (\citealt{Perley2019}).

In the optical counterpart spectrum of \EPb{}, we identified the emission lines associated with the underlying host galaxy, namely H$\alpha$ $\lambda$6564.6\footnote{ Rest wavelengths of the lines in vacuum are from https://classic.sdss.org/dr6/algorithms/linestable.html}, H$\beta$ $\lambda$4862.7, [O II] $\lambda$3728.5 and [O III] $\lambda$4960.3, 5008.2. We fitted multiple Gaussians to the emission line using the \textsc{lmfit}\footnote{https://lmfit.github.io/lmfit-py/} package and obtained the best-fit central wavelengths and their associated errors.  We infer a redshift of $z$= 0.457 $\pm$ 0.003 for \EPb{}.
\label{spec_section}

\begin{figure*}
\centering
\includegraphics[scale=0.45]{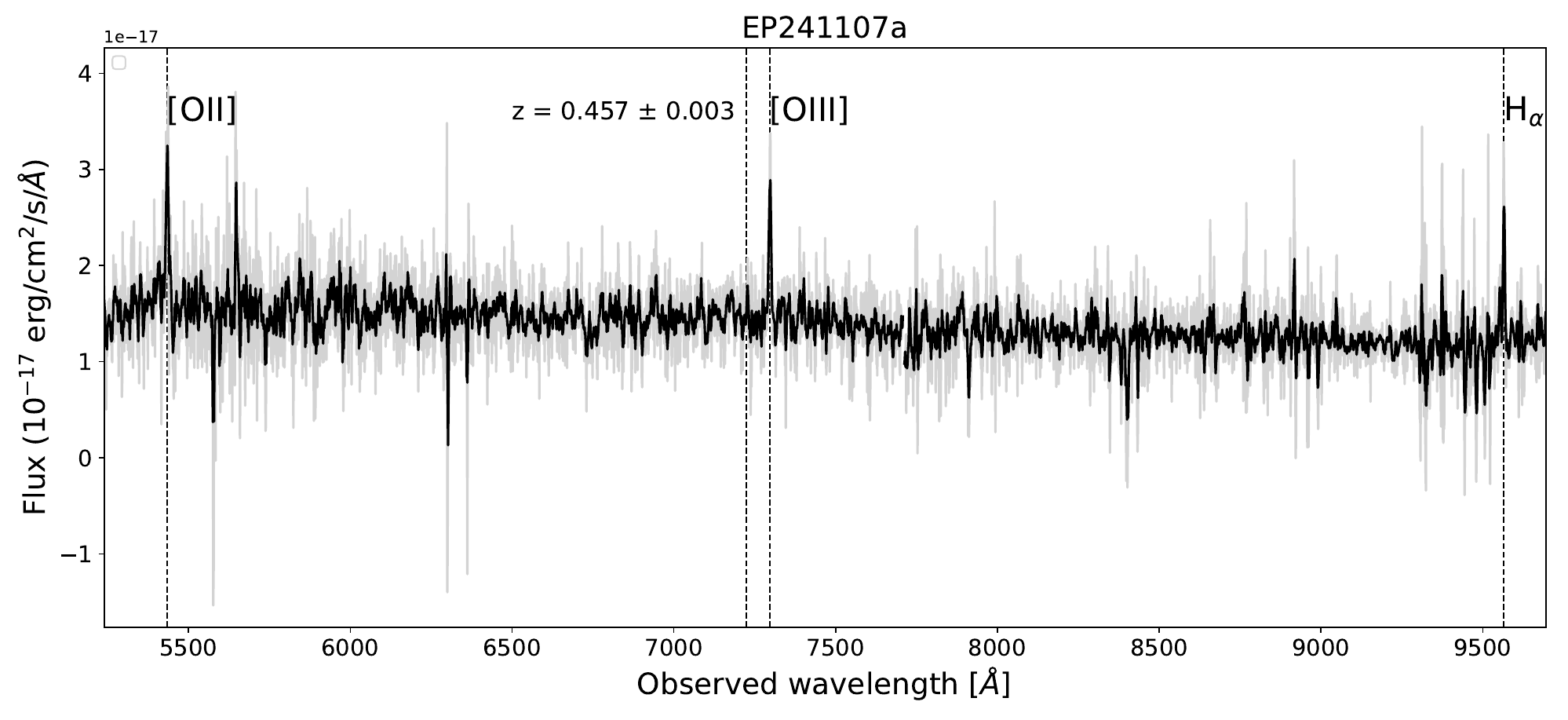}
\caption{  The flux calibrated Keck LRIS spectrum of the counterpart of \EPb{} taken 0.767 days after the EP-trigger is shown. For display purposes, the boxcar smoothed spectrum
(Box1DKernel with a width of 5 pixels) is shown in the black solid line, whereas the light grey shows the spectrum. The location of important emission lines in the spectrum is marked with dashed lines. The emission lines are the host contributions, and we derive a redshift of 0.457$\pm$0.003 from the emission lines.} 
\label{spec}
\end{figure*}

\subsection{Karl G. Jansky Very Large Array Radio Observations}

Under our approved Karl G Jansky Very Large Array (VLA) proposal 24B-492 (PI: Balasubramanian), we obtained observations of the field of \EPb{} in the  X band (central frequency of 10000\,MHz) and in the C band (central frequency of 6000\,MHZ) on 24 November 2024 and 01 December 2024 respectively. The raw data was calibrated using the automated \textsc{casa} VLA pipeline (version 6.5.4.9) and imaged along with self-calibration using an automated script \footnote{\url{https://science.nrao.edu/facilities/vla/data-processing/pipeline/VIPL654}}. We detect a point source-like counterpart to \EPb{} in both the frequency bands. Within 10$\arcsec$ of the radio detection, we did not find any existing radio source in the master radio catalogue \footnote{\url{https://heasarc.gsfc.nasa.gov/w3browse/master-catalog/radio.html}}.  This reaffirms that the radio detection is indeed associated with \EPb{}.  To obtain the flux density of the detected source, we use the \textsc{casa} task \textsc{imfit} and list the peak flux density and error values for the point source in Table \ref{tab}. The C band image is shown in Fig. \ref{detection_plot} ({\it right} panel).

\subsection{uGMRT Radio Observations}

We obtained upgraded Giant Metrewave Radio Telescope (uGMRT) observations of the field of \EPb{} in two frequency bands - band 4 (central frequency 750\, MHz, bandwidth 400\,MHz) and band 5 (central frequency 1260\,MHz, bandwidth 400\,MHz), on 20 December 2024 and 17 December 2024 respectively (ddtC404, PI: Eappachen). The raw data were downloaded in the \textsc{fits} format and converted to the \textsc{casa} \citep{2022PASP..134k4501C} measurement set format. Then the data was calibrated and imaged using the automated continuum imaging pipeline \textsc{casa-capture} \citep{2021ExA....51...95K}. Eight rounds of self-calibration were done within each pipeline run. We do not detect a significant radio source in the position of \EPb{}. We compute the 3\,$\sigma$ upper limits calculated from a large region centred on the source in the residual image and list the values in Table \ref{tab}.

\section{Modelling}

\subsection{Modelling the Gamma-Ray Burst Afterglow with {\sc Afterglowpy} }
\label{afterglowpy_method}

We fit the multiwavelength observations of \EPb{} using the publicly available semi-analytical package {\sc afterglowpy} (\citealt{2020ApJ...896..166R}) and {\sc emcee} (\citealt{2013PASP..125..306F}) code in \textsc{python}. The {\sc afterglowpy} code calculates GRB afterglow light curves and spectra (in Section \ref{discussion}, we discuss why we consider GRB as a viable model). It takes into consideration the effects of viewing angle and different jet structures. These structures describe how the energy of the jet changes with the jet opening angle $\theta_{c}$.

The {\sc afterglowpy} code allows the user to include viewing angle effects using the parameter $\theta_{\rm obs}$ and has an on-axis observation if $\theta_{\rm obs}$~=~0. {We assume a top-hat jet model for modelling the synchrotron emission from the jet interacting with the medium. } For a uniform jet, the set of free parameters is $\Theta$ = \{$E_0$, $\theta_c$, $\theta_{\rm obs}$ $n_0$, $p$, $\epsilon_e$, and $\epsilon_b$\} where $E_0$, the isotropic-equivalent kinetic energy,  $\theta_c$ is the half opening angle of the jet, $n_0$ is the density of the inter-stellar medium, $p$ is the power-law index of the electron energy distribution, and $\epsilon_e$ and $\epsilon_b$ are the fraction of thermal energy that is imparted to electrons and magnetic fields, respectively. We fixed the fraction of shock-accelerated electrons $\xi_N$ = 1 for our code. We fixed the luminosity distance $d_L$ based on our spectroscopic redshift (see Section \ref{spec_section}) and the adopted cosmological model mentioned in Section \ref{intro}.

A {\sc Scipy} \citep{2020SciPy-NMeth} \textsc{optimize} routine was used to obtain better guess parameters for the MCMC walkers. 100 walkers were initialised around this guess parameter set and allowed to explore the parameter space, in order to maximise the likelihood. Some walkers got stuck in low-probability regions and didn't allow the MCMC chains to converge. So, all walkers were re-initialised near the high probability region, allowing the MCMC chains to converge. The posterior values and corner plots to visualise the MCMC fit were computed using \textsc{ChainConsumer} \citep{Hinton2016}. 
The priors for our seven free parameters, the nature of the prior and the posterior values obtained are given in the Table~\ref{tab:priors}.

\begin{table}
    \centering
    \begin{tabular}{lccccc}
        \toprule
        Parameter  & Bounds & Units & Prior & Posterior \\
        \midrule
        log$_{10}$ ($E_{0})$   & (49, 54)  & erg  & logFlat & $51.16^{+0.07}_{-0.08}$ \\
        $\theta_{c}$  & (0, $\pi$/6)  & deg & Flat & $0.26^{+0.03}_{-0.02}$ \\
        $\theta_{\rm obs}$ & (0, $\pi$/6)  & deg & Flat & $0.16^{+0.02}_{-0.03}$ \\
        $p$ & (2, 3) & -- & Flat & $2.202^{+0.036}_{-0.039}$\\
        log$_{10}$ ($n_{0}$) & (-6,1) & cm$^{-3}$ & logFlat  & $-1.19^{+0.15}_{-0.10}$ \\ 
        log$_{10}$ ($\epsilon_e$) & (-4, log$_{10}$(1/3))   &  -- & logFlat  & $-0.70^{+0.10}_{-0.06}$ \\
        log$_{10}$ ($\epsilon_B$) & (-8, log$_{10}$(1/3))   &  -- & logFlat &  $>-0.97$\\
        \bottomrule
        \bottomrule
    \end{tabular}
    \caption{{Details of the jet afterglow model fitting}. 100 walkers were initialized either evenly spaced within the respective bounds or clustered around the initial value.}
    \label{tab:priors}
\end{table}

\subsection{Modelling of the Host of  \EPb{} using {\sc Bagpipes} }

We determine the host properties of  \EPb{} using {\sc Bagpipes} (Bayesian Analysis of Galaxies for Physical Inference and Parameter Estimation; \citealt{2018MNRAS.480.4379C}), incorporating the host magnitudes. Employing the {\sc MultiNest} sampling algorithm, {\sc Bagpipes} fits stellar population models to multiband photometric data, accounting for the star formation history and the transmission function of neutral and ionised interstellar medium (ISM) in broadband photometry and spectra.  

\begin{figure}
\includegraphics[scale=0.39 ]{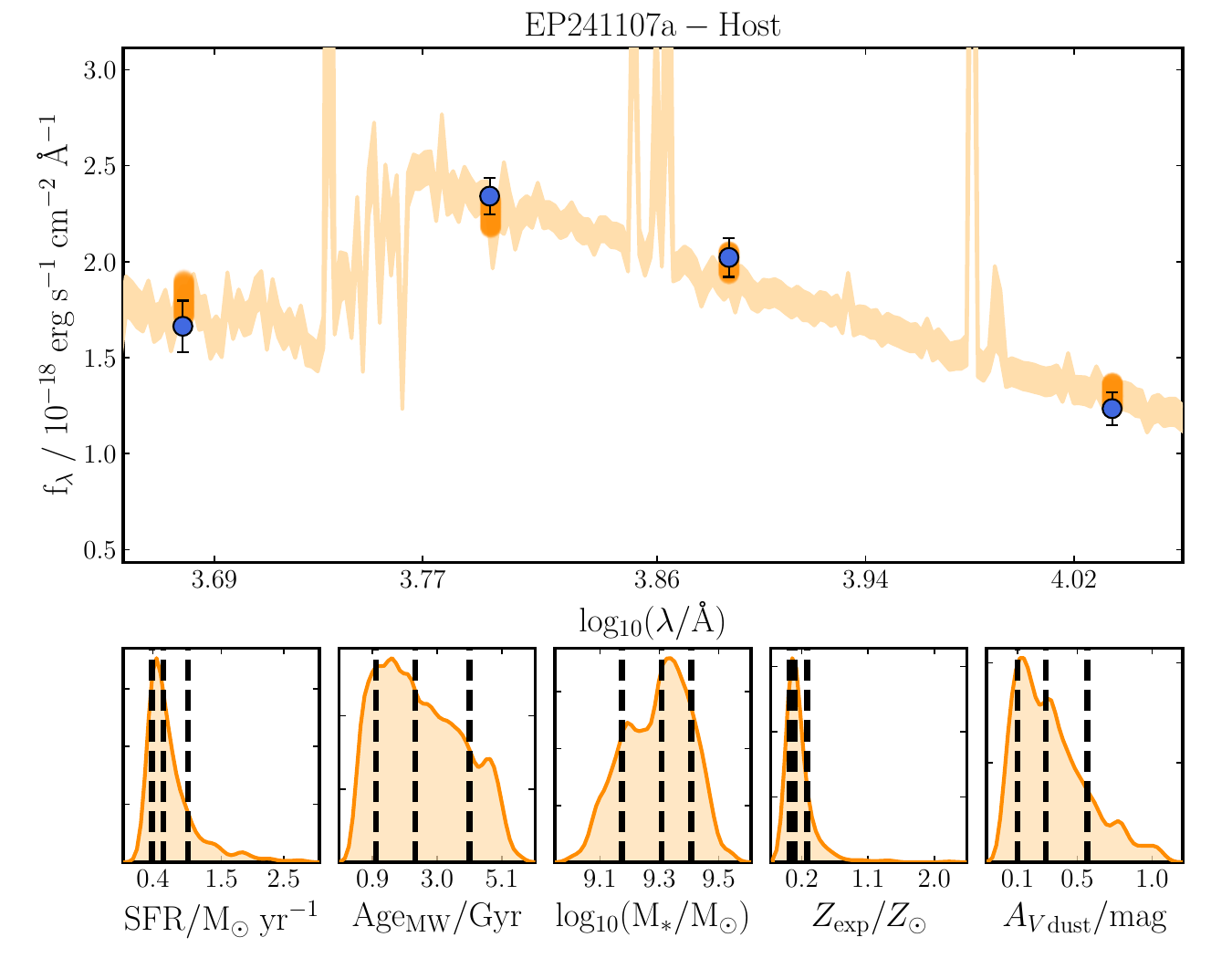}
\caption{ The  best-fitting SED model of the host of  \EPb{} using {\sc BAGPIPES} package is shown. The blue markers give the photometric data, and their 1$\sigma$ uncertainties, whereas the 16th–84th percentile range for the posterior probability for the spectrum and
photometry are shown in light and dark orange colours, respectively. The bottom panel shows the posterior probability distributions for the
ﬁve ﬁtted parameters, SFR, age, galaxy stellar mass, metallicity and dust extinction. In each subplot, the 16th, 50th, and 84th percentile posterior values are represented by vertical dashed black lines marked from left to right.}
\label{sed}
\end{figure}

The posterior probability distributions for the host galaxy age, dust extinction ($A_V$), star formation rate (SFR), metallicity ($Z$), stellar mass ($M_*$), and specific star formation rate (sSFR) are obtained through fitting with {\sc Bagpipes}, assuming a flat prior for the redshift based on the spectroscopic redshift. We adopt an exponentially declining star formation history function and apply the dust attenuation parameterisation developed by \citet{2000ApJ...533..682C} to model the spectral energy distributions (SEDs), using priors for $A_V$ in the range of 0.0 to 2.0 mag. The input observed photometric data are shown in blue.  The 16th to 84th percentile range of the posterior probability distribution for the spectrum and broadband photometry (shaded in light orange and orange) is presented in Fig. \ref{sed}. { We utilised the publicly available photometric magnitudes of the candidate host galaxy of \EPb{} from the DESI Legacy Imaging Survey Archive (\citealt{2019AJ....157..168D}). The reported magnitudes are $g = 23.64 \pm 0.08$, $r = 22.68 \pm 0.04$, $i = 22.39 \pm 0.05$, and $z = 22.19 \pm 0.07$.} The inferred host properties are discussed in Section \ref{host_prop}.

\section{Discussion}

\label{discussion}
\subsection{FXT Properties and Energetics}
We begin with the discussion on the X-ray observations of \EPb{}. \EPb{} showed a peak flux of approximately \(4.2 \times 10^{-9}~\mathrm{erg}~\mathrm{cm}^{-2}~\mathrm{s}^{-1}\) in the 0.5--4~keV band \citep{2024GCN.38171....1L}. Given its spectroscopic redshift of $\sim$ 0.457, the inferred X-ray luminosity at trigger is \(\sim 3.4 \times 10^{48}~\mathrm{erg}~\mathrm{s}^{-1}\) (in the 0.5--4~keV band by EP-WXT).
This luminosity is higher compared to the predicted X-ray luminosities of typical SN SBOs ($L_{\rm{X,peak}} \approxlt 10^{45}$ \lum{}  for supernova SBOs; \citealt{2008Natur.453..469S}; \citealt{2017hsn..book..967W}; \citealt{2022ApJ...933..164G}). Approximately 60 minutes post-trigger,  an X-ray afterglow with a luminosity of \(1.2 \times 10^{46}~\mathrm{erg}~\mathrm{s}^{-1}\)  was detected by EP-FoXT in the 0.5--10~keV band. This X-ray counterpart declined rapidly and went undetected $\sim$3.9 days post-trigger in the same energy band (\citealt{2024GCN.38171....1L}).
No $\gamma$-ray association has been reported so far for \EPb{}\footnote{At the time of writing} through GCNs. \citet{2025arXiv250605920Z} report the absence of gamma detection for \EPb{} from multi-instrument search.

\begin{figure*} 
\centering
\includegraphics[scale=0.85]{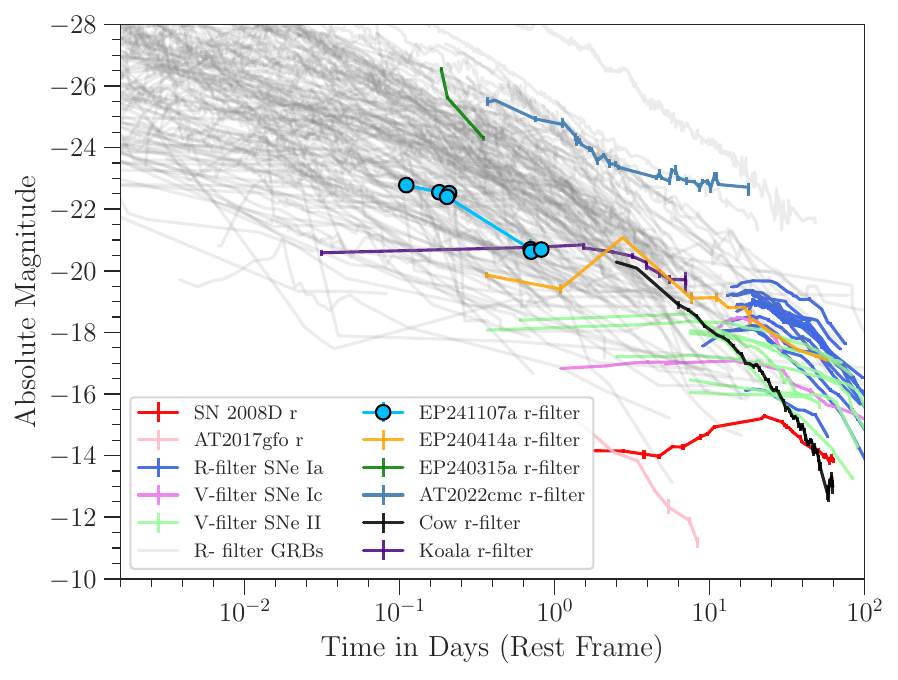}
\caption{The optical light curve of  \EPb{} ($r$-filter; shown in skyblue) in comparison with the light curves of SN2008D (\citealt{2008Natur.453..469S}), AT2017gfo (\citealt{2017ApJ...848L..17C}), Type II SNe, Type Ia SNe, Type Ic SNe (\citealt{2000ApJ...534..660I};  \citealt{2006AJ....131..527J}; \citealt{2006ApJ...645L..21M}; \citealt{2017ApJS..233....6H}), GRB afterglows (\citealt{2010ApJ...720.1513K}; \citealt{2011ApJ...734...96K}), the optical counterpart of EP240315a (\citealt{2024arXiv240416350L}; \citealt{2024ApJ...969L..14G}), EP240414a (\citealt{2024arXiv240919056V}; \citealt{2024arXiv240919070S}), AT2018cow (the ``Cow''; \citealt{2019MNRAS.484.1031P}), ZTF18abvkwla (the ``Koala''; \citealt{2020ApJ...895...49H}) and the jetted TDE AT2022cmc (\citealt{2022Natur.612..430A}). The magnitudes are corrected for the galactic extinction. The light curve of \EPb{} is consistent with the GRB afterglow light curves.}
\label{optical_comparison}
\end{figure*}

\subsection{Optical Evolution:  Comparison with Known Transient Classes}
In Fig. \ref{optical_comparison}, we compare the optical light curve of  \EPb{}, corrected for Galactic extinction,\footnote{Calculated using \url{https://ned.ipac.caltech.edu/extinction_calculator} { A$_V$ = 0.3, R$_V$ $\sim$ 3.1}}, with the light curves of SNe, GRB afterglows, other EP FXTs, and FBOTs.  We infer that the light curve \EPb{} is consistent with the population of GRB afterglows\footnote{For comparison, we consider the $R$ and $r$ filters to be similar.} (shown in grey in Fig. \ref{optical_comparison}). The $r$-filter light curves of EP240315a  and EP240414a \citep{2024ApJ...969L..14G, 2024arXiv240416350L, 2024arXiv240919056V, 2024arXiv240919070S} also fall within the parameter space occupied by the GRB population similar to \EPb{}. That said,  EP240414a was found to be associated with the core-collapse of a massive star, resulting in a broad line Type Ic SNe seen in lGRBs and its evolution does not resemble the light curve of \EPb{}. Furthermore, the association of  \EPb{} with AT2018cow-like transients appears unlikely based on a comparison of their optical light curves.

We observe a two-phase temporal evolution in the $r$-filter light curve of \EPb{}, with decay indices for the initial phase (prior to 0.3 days) to be $\alpha = 0.46 \pm 0.12$ and a later phase with an $\alpha = 1.33 \pm 0.13$. These decay indices are consistent with the optical afterglow light curves of GRBs (\citealt{2010ApJ...720.1513K}). The rapid optical decay is not consistent with the slow evolution typical of TDE  light curves in the early phase (\citealt{2020SSRv..216..124V}). Based on these optical light curve comparisons, we conclude that  \EPb{} is most likely associated with the GRB afterglow population, representing a new addition to the growing catalogue of EP-discovered GRBs.

\begin{figure*}
\centering
\includegraphics[scale=0.8]{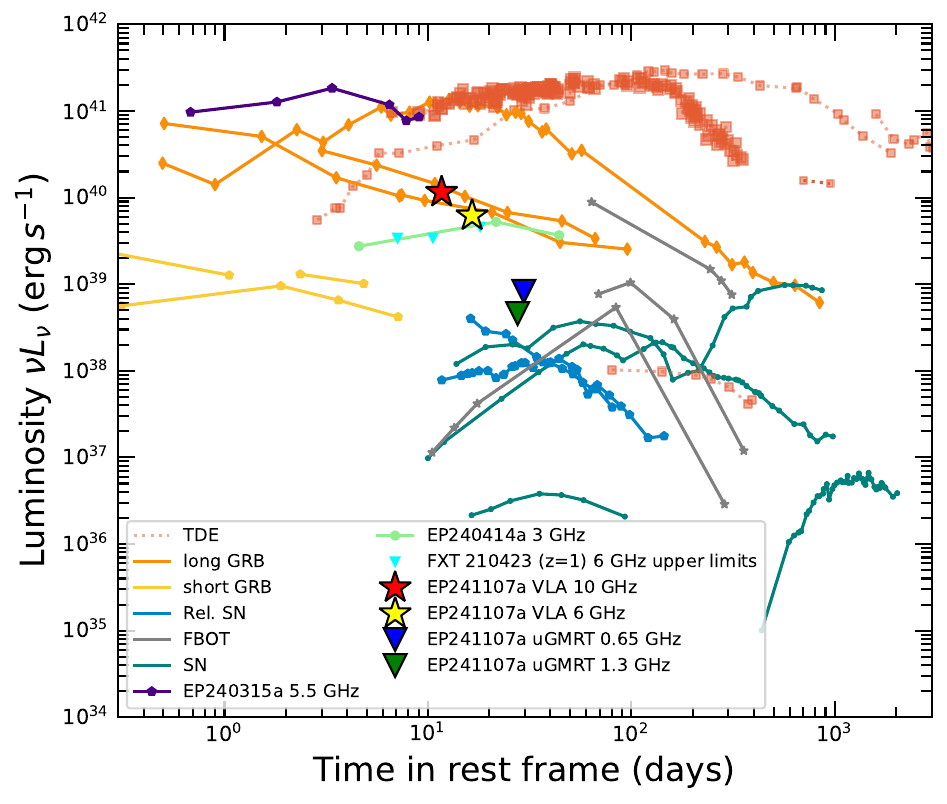}
\caption{The uGMRT and VLA radio observations of  \EPb{} compared with other extragalactic transients in radio, including relativistic supernovae, LFBOTs (\citealt{2020ApJ...895L..23C}), thermal TDEs (\citealt{2020SSRv..216...81A}),
relativistic TDEs (\citealt{2018ApJ...854...86E}; \citealt{2023MNRAS.521..389R}),
short gamma-ray bursts (\citealt{2021ApJ...906..127F}), long gamma-ray
bursts (\citealt{2003Natur.426..154B}; \citealt{2014MNRAS.444.3151V}; \citealt{2023ApJ...946L..23L}), and low-luminosity gamma-ray bursts (\citealt{1998Natur.395..663K, 2010Natur.463..513S}). The specific luminosity has been K-corrected to account for the cosmological distances of some sources. This plot is based on Fig. 9 in \citet{2020ApJ...895...49H} and references therein.}
\label{fig:radio_lums}
\end{figure*}

\subsection{Insights from the Radio Observations}

In Fig. \ref{fig:radio_lums}, we show 3$\sigma$ radio upper limits from our uGMRT observations and detections from VLA observations at 6 and 10 GHz. These limits and detections are compared with the radio light curves of other extragalactic transients in radio.  The specific luminosity has been K-corrected (\citealt{2002astro.ph.10394H}) to account for the cosmological distances of some sources. Additionally, we compare our upper limits with the radio light curves of previously reported EP-FXTs, namely EP240315a (\citealt{2024ApJ...969L..14G}; \citealt{2024arXiv240718311R}), EP240414a (\citealt{2024arXiv240919055B}), and the radio upper limits for the FXT XRT210423 (\citealt{2023ApJ...948...91E}), discovered from the {\it Chandra} archival data (\citealt{2024arXiv240707257I}). We infer that the VLA detections in both bands (See the star marker in Fig. \ref{fig:radio_lums}) are in agreement with the parameter space occupied by the radio afterglows of GRBs. We note that it is likely that the non-detection in the lower frequencies for \EPb{} might be because of absorption processes (\citealt{1998ApJ...499..810C}). In summary, the {radio} detection for \EPb{} strongly supports a GRB-like afterglow.

\subsection{Comparison of Host Properties}
\label{host_prop}

  We found a faint associated host centred at R.A. = $02^{\rm h}$$20^{\rm m}$$02.40^{\rm s}$, Dec = $+03^{\rm \circ}$ $20^{\rm \prime}$ $02.33^{\rm \prime \prime}$ which is $\sim$0.83$^{\rm \prime \prime}$  from the counterpart reported by \citet{GCN38115} in the DESI legacy images for \EPb{}. At the redshift of $\sim$0.457,  the projected offset is 4.8 kpc.  This galactocentric offset places \EPb{} at the $\sim$90th percentile of lGRB offset distribution (\citealt{2017MNRAS.467.1795L}), while it is also in agreement with the $\sim$40 percentile of sGRB offsets (\citealt{2022ApJ...940...56F}; \citealt{2022MNRAS.515.4890O}). 
 
\begin{figure} 
 \includegraphics[scale=0.54 ]{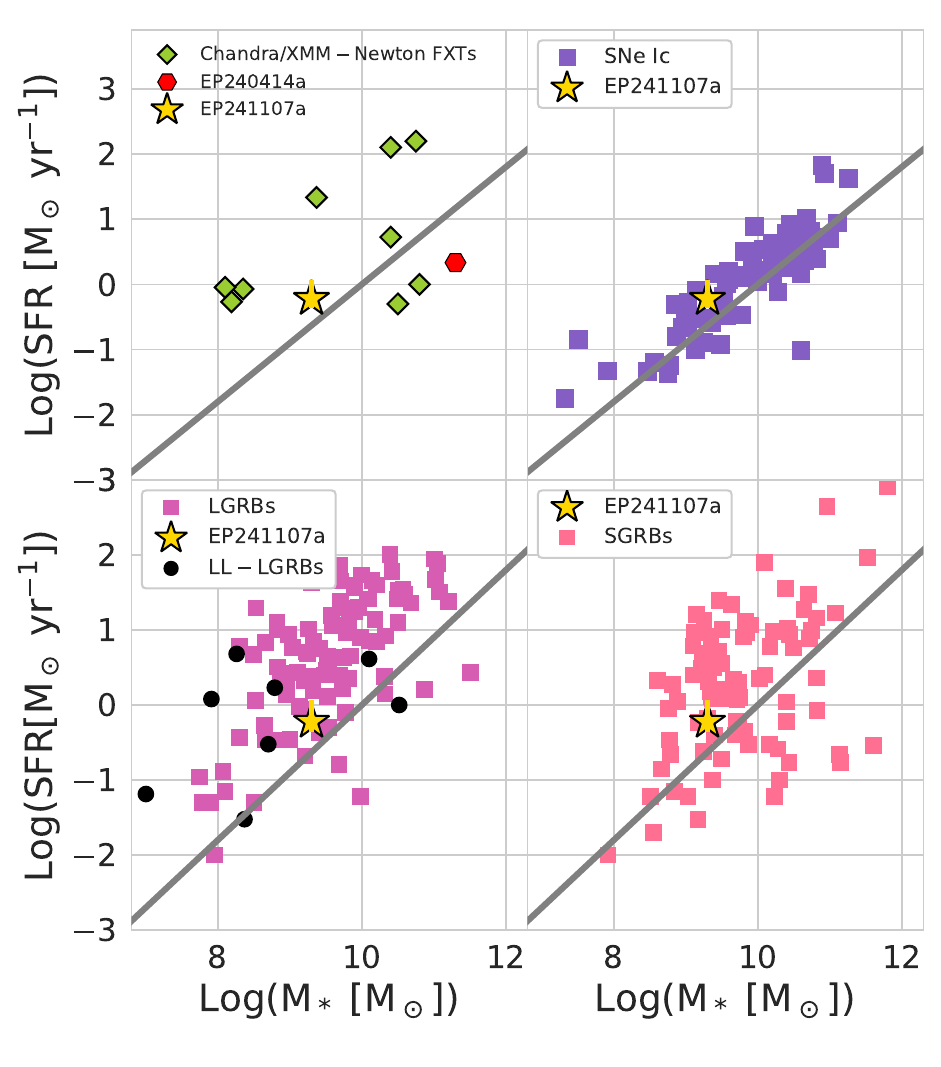}
\caption{ M$_{*}$ and SFR of the host galaxy of \EPb{} (yellow star) compared with host galaxies of other transient events.  Different panels show the host properties of the previously reported FXTs (\citealt{Quirola2022}; \citealt{Bauer}; \citealt{Xue2019}; \citealt{2023ApJ...948...91E},\citealt{2024MNRAS.52711823E}; \citealt{2024arXiv240919056V}), of lGRBs, sGRBs, low-luminosity lGRBs (\citealt{2007A&A...464..529W}; \citealt{2008A&A...490...45C}; \citealt{2014A&A...562A..70M}; \citealt{2014PASP..126....1L}; \citealt{2017A&A...602A..85K}; \citealt{2018ApJ...867..147W}; \citealt{2016ApJS..227....7L}; \citealt{2019MNRAS.485.5411A}; \citealt{2022ApJ...940...56F}), and SN Ic (\citealt{2021ApJS..255...29S}). The solid grey lines show the best-fit local galaxy main
sequence relation from \citet{2010ApJ...721..193P}. }
\label{host}
\end{figure}

 Using {\sc Bagpipes} and observed magnitudes of the candidate host in $g$, $r$, $i$ and $z$ filters from the legacy archive, we estimated the host properties for \EPb{}. A SFR of 0.6$\substack{+0.4 \\ -0.2}$ $\msun$yr$^{-1}$ and Log($M_{*}[\msun])$ of 9.3 $\pm$ 0.1 is estimated for the candidate host of \EPb{}. The Fig. \ref{host} shows the M$_{*}$ and SFR of the host galaxy of \EPb{} (gold star) compared with host galaxies of the previously reported FXTs (\citealt{Quirola2022}; \citealt{Bauer}; \citealt{2023ApJ...948...91E}, \citealt{2024MNRAS.52711823E}; \citealt{2024arXiv240919056V}; \citealt{2024arXiv241010015Q}), of lGRBs, sGRBs, low-luminosity lGRBs (\citealt{2007A&A...464..529W}; \citealt{2008A&A...490...45C}; \citealt{2014A&A...562A..70M}; \citealt{2014PASP..126....1L}; \citealt{2017A&A...602A..85K}; \citealt{2018ApJ...867..147W}; \citealt{2016ApJS..227....7L}; \citealt{2019MNRAS.485.5411A}; \citealt{2022ApJ...940...56F}), SN Ic (\citealt{2021ApJS..255...29S}). Though the M$_{*}$ and SFR of the host galaxy of \EPb{} are consistent with the populations of hosts of lGRBs and sGRBs, it takes up the parameter space previously not occupied by the {\it Chandra}-{\it XMM}-Newton FXTs.

It is worth noting that the host properties inferred for \EPb{} are based only on four observed magnitudes. Although our SED modelling provides an estimate of the host properties, deeper observations of the host galaxies would significantly improve our understanding of their environments and will be crucial in constraining the nature of the transients. A pertinent example is the case of CDF-XT2, where deeper JWST observations of the host galaxy by \citet{2024arXiv241010015Q} improved upon the earlier host properties reported by \citet{Xue2019}, thereby better constraining the nature of the transient.

\begin{figure}
\centering
\includegraphics[scale=0.9]{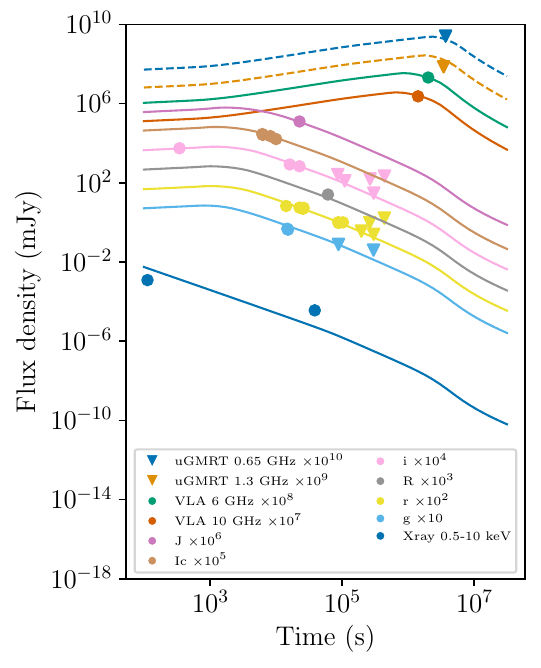}
\caption{ The broadband {\sc Afterglowpy} modelling of \EPb{} is presented,  where the figure shows the best-fit light curves. Each data point represents observed flux density in different bands ranging from X-ray to radio. Upper limits are indicated by inverted triangles. Note that the light curves are scaled for visual purposes. The best-fit jet parameters inferred from the {\sc Afterglowpy} modelling are discussed in detail in Section \ref{jet_parameters}.} 
\label{fig:afterglowpy}
\end{figure}

\subsection{Constraints on Relativistic Jets}
\label{jet_parameters}

 For \EPb{}, we discuss the jet parameters that we infer from our \textsc{afterglowpy} modelling discussed in Section \ref{afterglowpy_method}. We consider the simple uniform jet, which is sufficient to provide a good fit of the data set. The broadband \textsc{afterglowpy} light curves from radio to the X-ray band of \EPb{} are shown in Fig. \ref{fig:afterglowpy}. Note that the light curves are scaled for visual purposes. The corner plot of the posterior fit parameters is given in Appendix Fig. \ref{fig:cornerplot}. Our fit yields a jet of  $E_{\mathrm{K,iso}} \sim 10^{51}$ erg with a  $\theta_{c}$ $\approx$15$^{\circ}$ viewed at an angle of $\theta_{\rm obs}$ $\approx$9$^{\circ}$ which agrees with the population of cosmological GRBs. The best fit $n_0$ $\approx$ 10$^{-1.2}$ cm$^{-3}$, suggesting a typical low-density ISM environment for \EPb{} (\citealt{2012ApJ...746..156C}). The value of the power-law index of the electron energy distribution {\it p} from the best fit model $\sim$2.2, agrees with the value ({\it p} =2.3; \citealt{2005PhRvL..94k1102K}) inferred from the shock acceleration principle. We infer that in the case of 1.3 GHz observation, our upper limits do not agree with the best-fit light curve. One explanation could be that \textsc{afterglowpy} does not account for synchrotron self-absorption. The radio data show evidence for a jet break (for the 6 GHz light curve, we see a jet break around 10 days since the trigger). 
 
 Although our model favours an on-axis scenario ($\theta_{\rm obs}$ <  $\theta_{c}$), no gamma-ray detections were reported for \EPb{}\footnote{At the time of writing}. Considering the best fit $E_{\mathrm{K,iso}}$ $\sim$ 1.4 $\times$ 10$^{51}$ erg and the  $\theta_{c}$ $\approx$15$^{\circ}$, the beaming corrected kinetic energy $E_{K}$~=~(1 - cos $  \theta_{c}$) $E_{\mathrm{K,iso}}$ $\approx$ 4.8  $\times$ 10$^{49}$ erg. This beam corrected kinetic energy of 4.8  $\times$10$^ {49}$ erg, is at the lower end of the distribution of classical GRBs (\citealt{2018ApJ...859..160W}). Hence, the lack of gamma-rays could be interpreted as \EPb{} being an intrinsically low-energy GRB. Though the energy scales match with a few relativistic TDEs (e.g., XMMSL1 J0740-85; \citealt{2016ApJ...819L..25A}; \citealt{2017A&A...598A..29S}), we would expect the density of the medium to be three orders of magnitude greater in the case of TDEs (\citealt{2020NewAR..8901538D}).{ The optical counterpart of \EPb{} is roughly ten times fainter than the typical GRB afterglows one day after the trigger, when compared to the golden sample of \citet{2010ApJ...720.1513K}}. Therefore, while \EPb{} exhibits the energetics and jet geometry typical of cosmological GRBs, the absence of gamma-ray emission despite its inferred on-axis viewing geometry suggests that this event represents an intrinsically faint GRB at the lower end of the classical GRB energy distribution.

\subsection{ Evaluating \EPb{} in the Landscape of FXTs}

\begin{figure}
\centering
\includegraphics[scale=0.7]{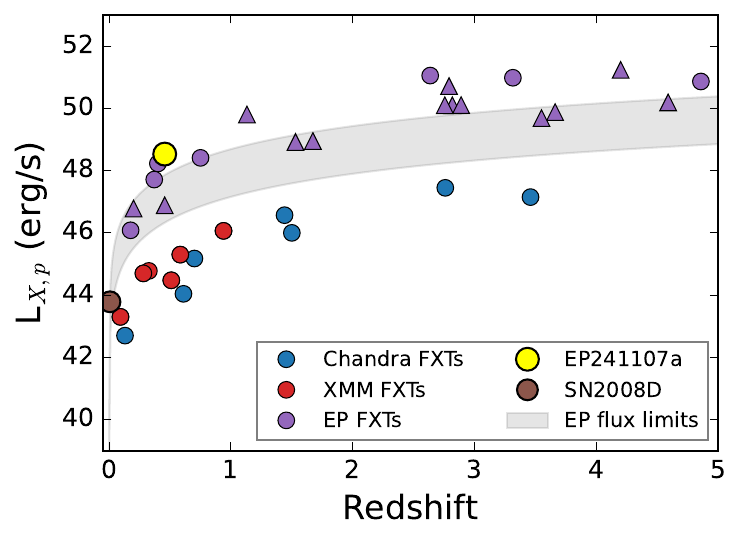}
\caption{ The peak X-ray luminosity of \EPb{} (yellow marker) is compared with those of extragalactic FXTs observed by {\it Chandra} (\citealt{Quirola2022}; \citealt{Quirola2023}), {\it XMM}-Newton (\citealt{Alp2020}; \citealt{2024MNRAS.52711823E}; \citealt{anneetal}), and EP (from GCNs reported by the EP team; \citealt{2022hxga.book...86Y}) as a function of redshift. The upward triangle shows the limit on the peak X-ray luminosity (often, in the case of EP-FXTs, the EP team reports the average unabsorbed flux). We also plot the sensitivity curve for EP (\citealt{2022hxga.book...86Y}). The details of pre-EP era FXTs are given in Table \ref{tab:FXT_full_properties}.}
\label{fig:Lx_z}
\end{figure}

In this section, we examine the nature of \EPb{} in relation to the broader FXT population detected by \textit{EP}, \textit{Chandra}, \textit{XMM}-Newton and {\it Swift}. In Fig.\ref{fig:Lx_z}, the peak X-ray luminosity of  \EPb{} (yellow) is compared with those of extragalactic FXTs observed by {\it Chandra} (\citealt{Quirola2022}; \citealt{2022ApJ...927..211L}; \citealt{Quirola2023}; \citealt{2023ApJ...948...91E}; \citealt{2024arXiv240707257I}), {\it XMM}-Newton (\citealt{Alp2020}; \citealt{2024MNRAS.52711823E}; \citealt{anneetal}), {\it Swift} (\citealt{2008Natur.453..469S}) and EP (from GCNs reported by EP team; \citealt{2022hxga.book...86Y}) as a function of redshift (similar to the figure from \citealt{2024arXiv240919070S}).  The limit on the peak X-ray luminosity (often in the case of EP-FXTs, the EP team reports the average unabsorbed flux) is shown with an upward triangle. The sensitivity curve for EP is also plotted in the Fig. \ref{fig:Lx_z}.

 The extragalactic FXTs discovered from {\it Chandra} and {\it XMM}-Newton show peak X-ray luminosities ranging from $L_{X,p} \sim 10^{42}$--$10^{47}$~\lum, at redshifts $z \sim 0.09$--3.5. The most distant FXT among the pre-EP era is CDF-XT2, which is at a redshift of $\sim$3.5 with a peak X-ray luminosity of $\approx$ 10$^{47}$ \lum (\citealt{2024arXiv241010015Q}). 

On the other hand, EP-discovered FXTs exhibit peak X-ray luminosities of $L_{X,p} > 10^{46}\,\mathrm{erg\,s^{-1}}$ and span a redshift range of $z \sim 0.176$--4.859, with a mean redshift of $z \sim 2$. The nearest EP-detected FXT with a spectroscopic redshift is EP250108a at $z = 0.176$ \citep{2025arXiv250408889R,2025arXiv250408886E,2025arXiv250417034L,2025arXiv250417516S}, which was associated with the closest known broad-lined Type Ic SN discovered by EP; meanwhile, the most distant case is of EP240315a \citep{2024ApJ...969L..14G,2024arXiv240416350L,2025NatAs...9..564L}.

FXTs exhibit diverse properties and are believed to originate from a variety of progenitor mechanisms. For example, CDF--XT2, discovered in the {\it Chandra} archive, may be associated with a low-luminosity collapsar progenitor (\citealt{Xue2019}; \citealt{2024arXiv241010015Q}), while FXT XRT~110621, found through a search of archival {\it XMM}-Newton data, could plausibly be an SN SBO (\citealt{Alp2020}; \citealt{2024MNRAS.52711823E}). Additionally, FXTs such as XRT~210423 (\citealt{2022MNRAS.514..302E}; \citealt{2024arXiv240707257I}) may be linked to a BNS merger scenario, in which the central engine is powered by a millisecond magnetar (\citealt{2014MNRAS.439.3916M}; \citealt{2017ApJ...835....7S}; \citealt{2025arXiv250501606C}). However, \citet{2025arXiv250611676B} considered all plausible formation pathways for millisecond magnetars that could produce an FXT and inferred that such progenitors can account for at most 10 per cent of the entire FXT population. {Also, a TDE origin remains a plausible explanation for FXTs like XRT~000519} (\citealt{2013jonker}; \citealt{2022MNRAS.514..302E}). As seen in the Fig. \ref{fig:Lx_z}, the variety of FXT progenitor scenarios is reflected in the wide range of X-ray luminosities and associated redshifts.  \citet{2024arXiv240919070S} point out that the X-ray luminosity parameter space explored by EP is distinct from that of previously discovered FXTs by \textit{Chandra} and \textit{XMM}-Newton, owing to the different observing strategy, probing high redshift and intrinsically more luminous events.

A large number of EP-discovered FXTs have been associated with GRBs or show similar properties to them (for e.g., \citealt{2024arXiv240416350L}). 
\EPb{} properties also agree with the typical afterglow nature seen in GRBs. \EPb{} is consistent with the population of FXTs detected by EP and are mostly distinct from the population of FXTs detected by  {\it Chandra} and {\it XMM}-Newton. CDF-XT1 and CDF-XT2 are exceptions as they are associated with host galaxies at $z$>2.5. However, at least for some of the {\it Chandra} and {\it XMM}-Newton FXTs, the energetics might vary with future deeper observations of the host, as some might be wrongly associated with the host at a different redshift (for e.g., see \citealt{2024arXiv241010015Q}).

\section{Conclusion}
We present the multiwavelength follow-up of an EP-detected FXT  \EPb{}. A radio counterpart was detected at 10 GHz and 6 GHz with flux densities of 232~$\pm$~9 $\mu$Jy and 207~$\pm$~8 $\mu$Jy, respectively, between $\sim$16-- 24 days after the X-ray trigger. We compared the radio and optical light curves of \EPb{} with those of other transients and found that \EPb{} is consistent with the parameter space occupied by GRB afterglows.

We also examined the host properties of \EPb{} and the host offsets, SFR and $M_{*}$, which match with the population of GRBs. We modelled the multiwavelength afterglow for \EPb{} and inferred the jet properties.  Our fit yields a jet of  $E_{\mathrm{K,iso}} \sim 10^{51}$ erg with a  $\theta_{c}$~$\approx$15$^{\circ}$ viewed at an  angle of $\theta_{\rm obs}$ $\approx$9$^{\circ}$. The absence of gamma-ray emission, the inferred on-axis viewing geometry and the energetics derived from afterglow modelling suggest that \EPb{} is an intrinsically faint GRB.

Finally, we compared the population of EP-FXTs with those detected by the \textit{Chandra}, \textit{XMM}-Newton, and {\it Swift} detected FXTs from the pre-EP era. \EPb{} is consistent with the population of FXTs detected by EP and are mostly distinct from the population of FXTs detected by  {\it Chandra} and {\it XMM}-Newton. However, at least for some of the pre-EP era FXTs, the energetics may change with future deeper observations of the host, as some might be wrongly associated with a host at a different redshift. The EP mission is detecting a significant number of FXTs, which may be associated with diverse progenitor mechanisms, highlighting the importance of prompt and multiwavelength follow-up observations to characterise and better understand the nature of these EP-FXTs.

\section{Acknowledgements}
{We thank the anonymous referee for the helpful comments}.
DE thank Jonathan Quirola-Vásquez,  Ashley Chrimes, Judhajeet Basu, and Hrishav Das for the useful discussions. DE acknowledges Lauren Rhodes for sharing the radio data of transients. AB and DE also thank Suvedha Suresh Naik for the discussions on MCMC. DE acknowledges the use of AI language models (Claude by Anthropic and ChatGPT by OpenAI) for assistance with code editing, data visualisation, spell checking, and grammar correction. GCA acknowledges support from the Indian National Science Academy (INSA) through its Senior Scientist Programme.

We thank the staff of the uGMRT that made these observations possible. uGMRT is run by the National Centre for Radio Astrophysics of the Tata Institute of Fundamental Research. The uGMRT observations of \EPb{} were carried out under  ddtC404, with D. Eappachen as PI.

The National Radio Astronomy Observatory is a facility of the National Science Foundation operated under cooperative agreement by Associated Universities, Inc.  Our \EPb{} Very Large Array (VLA) observations were carried out under the DDT proposal
24B-492 (PI: Balasubramanian). 

This work is partially based on data obtained with
the 2m Himalayan Chandra Telescope of the Indian Astronomical Observatory (IAO) under the proposal HCT-2024-C3-P32 (PI: D. Eappachen). We thank the staff of IAO,
Hanle, and CREST, Hosakote, that made these observations possible. The facilities at IAO and CREST are
operated by the Indian Institute of Astrophysics, Bangalore.

The GROWTH India Telescope (GIT) is a 70-cm telescope with a 0.7-degree field of view, set up by the Indian Institute of Astrophysics (IIA) and the Indian Institute of Technology Bombay (IITB) with funding from  Indo-US Science and Technology Forum and the Science and Engineering Research Board, Department of Science and Technology, Government of India. It is located at the Indian Astronomical Observatory (Hanle), operated by IIA. We acknowledge funding by the IITB alumni batch of 1994, which partially supports the operations of the telescope.

Some of the data presented herein were obtained at
Keck Observatory, which is a private 501(c)3 non-profit
organization operated as a scientific partnership among
the California Institute of Technology, the University
of California, and the National Aeronautics and Space
Administration. The Observatory was made possible by
the generous financial support of the W. M. Keck Foundation. The authors wish to recognize and acknowledge
the very significant cultural role and reverence that the
summit of Maunakea has always had within the Native Hawaiian community. We are most fortunate to
have the opportunity to conduct observations from this
mountain.

Based on observations obtained at the Southern Astrophysical Research (SOAR) telescope, which is a joint project of the Minist\'{e}rio da Ci\^{e}ncia, Tecnologia e Inova\c{c}\~{o}es (MCTI/LNA) do Brasil, the US National Science Foundation’s NOIRLab, the University of North Carolina at Chapel Hill (UNC), and Michigan State University (MSU).


\section*{Data Availability}
The data relevant to this article will be shared on request to the corresponding author.



\bibliographystyle{mnras}
\bibliography{example} 




\appendix
\section{Additional Figures and Tables}

\begin{figure*}
    \centering
    \includegraphics[scale=0.40]{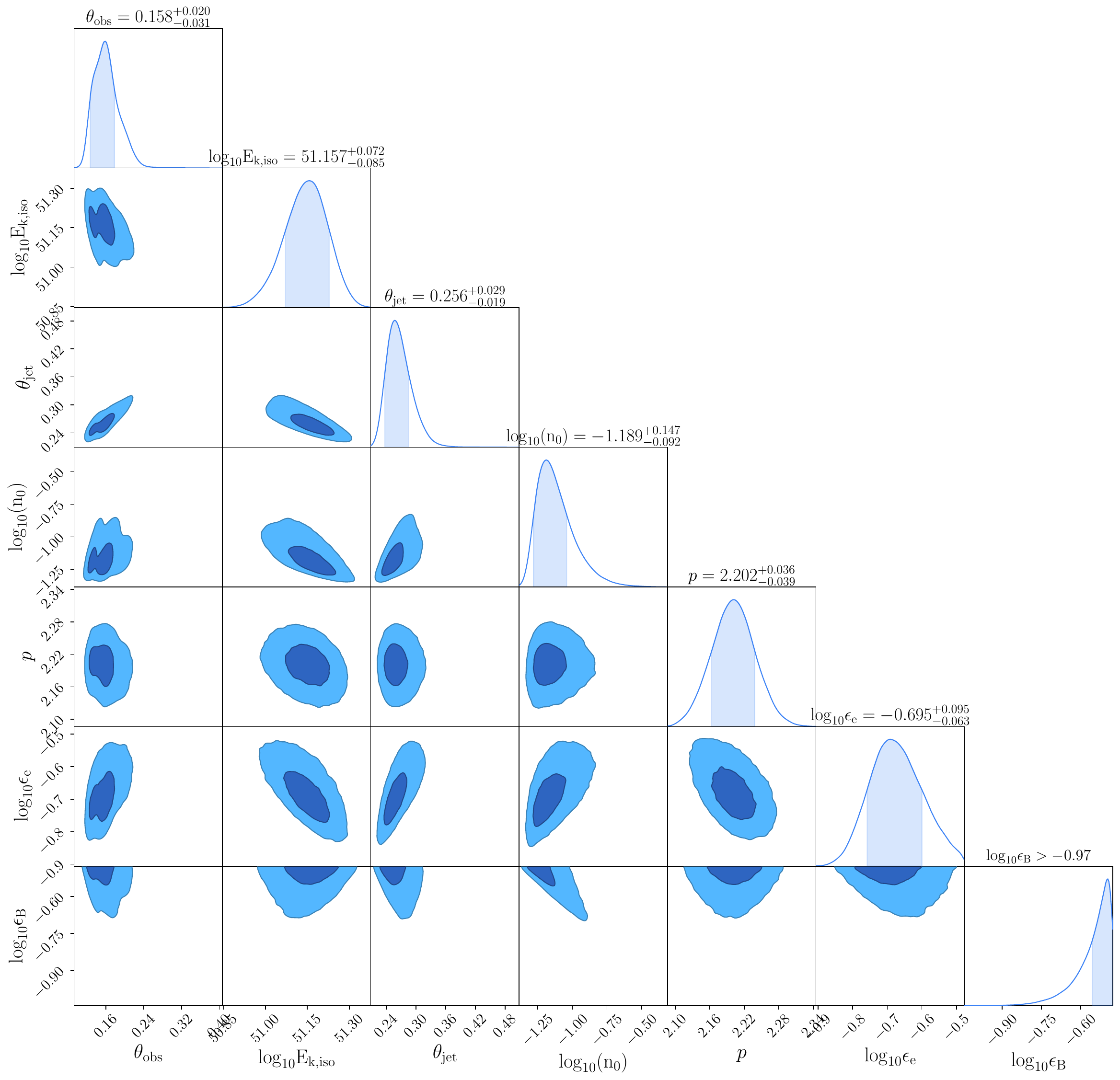}
    \caption{Corner plot for \EPb{}}
    \label{fig:cornerplot}
\end{figure*}

\begin{table*}
\centering
\caption{Details of the FXTs used in Fig \ref{fig:Lx_z} from the pre-EP era.}
\begin{tabular}{|l|c|c|c|c|l|}
\hline
FXT & $L_{x,p}$ (erg\,s$^{-1}$) & $z_\mathrm{phot}$ & $z_\mathrm{spec}$ & Telescope & Reference \\
\hline
XRT~030206  & \(5 \times 10^{44}\)   & --   & \(0.281 \pm 0.003\)  & \textit{XMM}-Newton & \citet{2024MNRAS.52711823E} \\
XRT~041230  & \(1.1 \times 10^{44}\) & \(0.61^{+0.13}_{-0.17}\) & --  & \textit{Chandra} & \citet{Quirola2022} \\
XRT~060207$^*$  & 1.2 $\times$ 10$^{46}$ & --  & 0.939  & {\it XMM}-Newton & \citet{anneetal} \\
XRT~080819  & \(1.5 \times 10^{45}\) & 0.7$^{+0.04}_{-0.10}$
   & --   & \textit{Chandra} & \citet{Quirola2022} \\
XRT~100424$^{\diamond}$   & \(5 \times 10^{42}\)   & 0.13  & --   & \textit{XMM}-Newton & \citet{Alp2020} \\
XRT~110621  & \(2 \times 10^{43}\)   & --   & 0.0928 $\pm$ 0.0002 & \textit{XMM}-Newton & \citet{2024MNRAS.52711823E} \\
XRT~151128 & 3 $\times$ 10$^{44}$ & 0.51 $\pm$ 0.01 & -- & \textit{XMM}-Newton & \citet{2024MNRAS.52711823E} \\
XRT~151219 & \(2 \times 10^{45}\) & -- & 0.584 $\pm$ 0.009 & \textit{XMM}-Newton  & \citet{2024MNRAS.52711823E} \\
XRT~170901 & 3.7 $\times$ 10$^{46}$ & 1.44 $\pm$ 0.08 & -- & \textit{Chandra} & \citet{Quirola2023}; \citet{2022ApJ...927..211L}\\
XRT~141001/CDF XT1 &  2.8 $\times$ 10$^{47}$ & 2.76$^{+0.21}_{-0.13}$ & -- & \textit{Chandra} & \citet{2024arXiv241010015Q} \\
XRT~150322/CDF XT2 & 1.4 $\times$ 10$^{47}$ & -- & 3.4598 $\pm$ 0.0022 & \textit{Chandra} & \citet{2024arXiv241010015Q} \\
XRT~210423$^a$  & 1 $\times$ 10$^{46}$ & -- & 1.5082 $\pm$ 0.0001 &  \textit{Chandra} & \citet{2023ApJ...948...91E}; \citet{2024arXiv240707257I}\\
XRO~080109/SN2008D & 6.1 $\times$ 10$^{43}$ & -- & 0.00649 ($d$=27 Mpc) & \textit{Swift}-XRT & \citet{2008Natur.453..469S}\\
\hline
\end{tabular}
\\
$^*$ Considering the galaxy within the $1\sigma$ X-ray uncertainty region as the host. \\
$^{\diamond}$  Based on the combined information of three galaxies/average $z$; see \citet{Alp2020}. \\
$^a$ Assuming cNE is the host galaxy of XRT~210423 (\citealt{2023ApJ...948...91E}; \citealt{2024arXiv240707257I} )
\label{tab:FXT_full_properties}

\end{table*}


\label{lastpage}
\end{document}